\documentstyle[12pt]{article}
\begin{document}
\begin{center}
{\noindent{\bf{Covariant and Contravariant Vectors }}}

\vspace{0.5cm}

Alok Kumar{\footnote{e-mail address: alok@iiserbhopal.ac.in}} \\
IISER, Bhopal\\
ITI Campus (Gas Rahat) Building\\
Govindpura, Bhopal - 23 \\
India. \\
\end{center}

\vspace{3.5cm}            

{\noindent{\it{Abstract}}}

Vector is a physical quantity and it does not depend on any co-ordinate system. It need to be expanded in some basis for practical calculation and its components do depend on the chosen basis. The expansion in orthonormal basis is mathematically simple. But in many physical situations we  have  to choose an non-orthogonal basis (or oblique co-ordinate system). But the expansion of a vector in non-orthogonal basis is  not convenient to work with. With the notion of contravariant and covariant components of a vector, we make non-orthogonal basis to behave like orthonormal basis. The same notion appears in quantum mechanics as Ket and Bra vectors and we compare  the two equivalent situation via the completeness relation. This notion appears in the differential geometry of a metric manifold for tangent vectors at a point,  where it takes into account the non-orthogonality of basis as well as non-Euclidean geometry. 
 
\vspace{0.5cm}

\newpage 

Any arbitrary  vector $\vec{V}$ can be expanded in terms of  the orthonormal basis  $\hat{i},\hat{j},$ and $\hat{k} $ as [1]
\begin{equation}
\vec{V} = (\vec{V}.\hat{i})\hat{i}+(\vec{V}.\hat{j})\hat{j}+(\vec{V}.\hat{k})\hat{k}.
\end{equation}
The expansion of  equation (1) is the manifestation of  the completeness relation for orthonormal basis which in the dyadic form 
\begin{equation}
I = \hat{i}\hat{i}+\hat{j}\hat{j}+\hat{k}\hat{k}.
\end{equation}
The equation (1)  can be written in the matrix form 
\begin{eqnarray}
\vec{V}=\left[
\begin{array}{ccc}
\vec{V}.\hat{i}&0&0\\
0&\vec{V}.\hat{j}&0\\
0&0&\vec{V}.\hat{k}
\end{array}\right]\left[\begin{array}{c}
\hat{i}\\
\hat{j}\\
\hat{k}
\end{array}\right]
\end{eqnarray}
and there are no off-diagonal elements in the matrix of equation (3), for  the orthonormal basis. Thus in  the orthogonal basis, any arbitrary vector $\vec{V}$ is represented by a diagonal matrix and it is a mathematical privilege.

\vspace{0.5cm}

In the non-orthogonal basis ($\vec{a}$, $\vec{b}$, $\vec{c}$) any arbitrary vector  $\vec{V}$ can be expanded as
\begin{equation}
\vec{V} = V_{x}\vec{a}+V_{y}\vec{b}+V_{z}\vec{c}.
\end{equation}
We require  to determine $V_{x}$, $V_{y}$ and $V_{z}$ in terms of given vector $\vec{V}$ and  non-orthogonal basis vectors $\vec{a}$, $\vec{b}$ and $\vec{c}$, in order to compare non-orthogonal basis expansion  with the orthogonal basis expansion of  the equation (1). We take dot product both sides of equation (4) with $\vec{a}$, $\vec{b}$ and $\vec{c}$
\begin{equation}
\vec{V}.\vec{a} = V_{x} + V_{y} \vec{b}.\vec{a} + V_{z} \vec{c}.\vec{a},
\end{equation}
\begin{equation}
\vec{V}.\vec{b} = V_{x}\vec{a}.\vec{b} + V_{y}  + V_{z} \vec{c}.\vec{b},
\end{equation}
\begin{equation}
\vec{V}.\vec{c} = V_{x}\vec{a}.\vec{c} + V_{y} \vec{b}.\vec{c} + V_{z} .
\end{equation}
Equations (5), (6), (7) combined into the matrix equation
\begin{eqnarray}
\left[\begin{array}{c}
\vec{V}.\vec{a}\\
\vec{V}.\vec{b}\\
\vec{V}.\vec{c}\end{array}\right]=\left[
\begin{array}{ccc}
1 &\vec{a}.\vec{b}&\vec{a}.\vec{c}\\
\vec{a}.\vec{b}&1&\vec{b}.\vec{c}\\
\vec{a}.\vec{c}&\vec{b}.\vec{c}&1\\
\end{array}\right]\left[\begin{array}{c}
V_{x}\\
V_{y}\\
V_{z}
\end{array}\right].
\end{eqnarray}
The matrix equation (8) is  written 
\begin{eqnarray}
\left[\begin{array}{c}
V_{x}\\
V_{y}\\
V_{z}
\end{array}\right]
=\left[
\begin{array}{ccc}
1 &\vec{a}.\vec{b}&\vec{a}.\vec{c}\\
\vec{a}.\vec{b}&1&\vec{b}.\vec{c}\\
\vec{a}.\vec{c}&\vec{b}.\vec{c}&1\\
\end{array}\right]^{-1}\left[\begin{array}{c}
\vec{V}.\vec{a}\\
\vec{V}.\vec{b}\\
\vec{V}.\vec{c}\end{array}\right]
\end{eqnarray}
with the condition on determinant $D$
\begin{equation}
D= 1+2(\vec{a}.\vec{b})(\vec{a}.\vec{c})(\vec{b}.\vec{c})-(\vec{a}.\vec{c})^2-(\vec{a}.\vec{b})^2-(\vec{b}.\vec{c})^2 \ne 0.
\end{equation}
The solution of matrix equation (8) 
\begin{equation}
\left.
\begin{array}{ccc}
V_{x}=[A_{11}(\vec{V}.\vec{a})+A_{12}(\vec{V}.\vec{b})+A_{13}(\vec{V}.\vec{c})]\\
V_{y}=[A_{21}(\vec{V}.\vec{a})+A_{22}(\vec{V}.\vec{b})+A_{23}(\vec{V}.\vec{c})]\\
V_{z}=[A_{31}(\vec{V}.\vec{a})+A_{32}(\vec{V}.\vec{b})+A_{33}(\vec{V}.\vec{c})]\\
\end{array}
\right\}
\end{equation}
where all $A_{ij}$ are function  of scalars made of  vectors $\vec{a}$, $\vec{b}$ and  $\vec{c}$ and given in the Appendix. Equation (8) with solutions of  $V_{x}$, $V_{y}$, and  $V_{z}$ as in equation (11) is given by
\begin{eqnarray}
\vec{V}.\left[\begin{array}{c}
\vec{a}\\
\vec{b}\\
\vec{c}\end{array}\right]=\vec{V}.\left[
\begin{array}{ccc}
1 &\vec{a}.\vec{b}&\vec{a}.\vec{c}\\
\vec{a}.\vec{b}&1&\vec{b}.\vec{c}\\
\vec{a}.\vec{c}&\vec{b}.\vec{c}&1\\
\end{array}\right]\left[\begin{array}{c}
A_{11}\vec{a}+A_{12}\vec{b}+A_{13}\vec{c}\\
A_{21}\vec{a}+A_{22}\vec{b}+A_{23}\vec{c}\\
A_{31}\vec{a}+A_{32}\vec{b}+A_{33}\vec{c}\\
\end{array}\right].
\end{eqnarray}
The matrix equation (12) is true for any arbitrary vector $\vec{V}$ and hence it follows
\begin{eqnarray}
I=\left[
\begin{array}{ccc}
1 &\vec{a}.\vec{b}&\vec{a}.\vec{c}\\
\vec{a}.\vec{b}&1&\vec{b}.\vec{c}\\
\vec{a}.\vec{c}&\vec{b}.\vec{c}&1\\
\end{array}\right]\left[\begin{array}{c}
A_{11}\vec{a}+A_{12}\vec{b}+A_{13}\vec{c}\\
A_{21}\vec{a}+A_{22}\vec{b}+A_{23}\vec{c}\\
A_{31}\vec{a}+A_{32}\vec{b}+A_{33}\vec{c}\\
\end{array}\right]
\left[\begin{array}{c}
\vec{a}\\
\vec{b}\\
\vec{c}\\
\end{array}\right]^{-1}.
\end{eqnarray}
The matrix equation (13) is the completeness relation for  non-orthogonal basis ($\vec{a}$, $\vec{b}$, $\vec{c}$). In comparison to equation (2) this completeness relation for  non-orthogonal basis is not easy to work with. In many physical situations we are with the non-orthogonal basis to work. Therefore, it is natural to find some notion to handle the non-orthogonal basis. With the notion of contravariant and covariant vectors, we make  non-orthogonal basis to behave like orthogonal basis. This is the main purpose of this article to explore the notion of contravariant and covariant vectors in different areas of physics.

For a given non-orthogonal basis ($\vec{a}$, $\vec{b}$, $\vec{c}$), we can always construct an unique non-orthogonal basis ($\vec{a^{\prime}}$, $\vec{b^{\prime}}$, $\vec{c^{\prime}}$)[1,2] 
\begin{eqnarray}
\vec{a^{\prime}}=\frac{(\vec{b}\times\vec{c})}{\vec{a}.(\vec{b}\times\vec{c})};\
\vec{b^{\prime}}=\frac{(\vec{c}\times\vec{a})}{\vec{a}.(\vec{b}\times\vec{c})};\
\hat{c^{\prime}}=\frac{(\vec{a}\times\vec{b})}{\vec{a}.(\vec{b}\times\vec{c})};\
\end{eqnarray}
where $\vec{a}.(\vec{b}\times\vec{c})$ is the scalar triple product. The unique vectors ( $\vec{a'}$, $\vec{b'}$, $\vec{c'}$) given by equation (14) are called reciprocal set of vectors for a given set of vectors ($\vec{a}$, $\vec{b}$, $\vec{c}$) and it  satisfy the following properties [1,2]:
\begin{eqnarray}
\vec{a}.\vec{a^{\prime}}=\vec{b}.\vec{b^{\prime}}=\vec{c}.\vec{c^{\prime}}=1;\
\vec{a}.\vec{b^{\prime}}=\vec{a}.\vec{c^{\prime}}=0;\
\vec{b}.\vec{a^{\prime}}=\vec{b}.\vec{c^{\prime}}=0;\
\vec{c}.\vec{a^{\prime}}=\vec{c}.\vec{b^{\prime}}=0
\end{eqnarray}
The reciprocal vectors of ($\vec{a^{\prime}}$, $\vec{b^{\prime}}$, $\vec{c^{\prime}}$) are   ($\vec{a}$, $\vec{b}$, $\vec{c}$) and two set of vectors  ($\vec{a'}$, $\vec{b'}$, $\vec{c'}$) and ($\vec{a}$, $\vec{b}$, $\vec{c}$) are dual to each other. The only right-handed self-reciprocal (self-dual) sets of vectors are orthonormal triad $\hat{i}, \hat{j}, \hat{k}$ [1].

Dot product of equation (4) with vectors $\vec{a^{\prime}}$, $\vec{b^{\prime}}$ and $\vec{c^{\prime}}$, equation (9) reads  
\begin{eqnarray}
\left[\begin{array}{c}
V_{x}\\
V_{y}\\
V_{z}
\end{array}\right]
=\left[
\begin{array}{ccc}
1 &\vec{a^{\prime}}.\vec{b}&\vec{a^{\prime}}.\vec{c}\\
\vec{a}.\vec{b^{\prime}}&1&\vec{b^{\prime}}.\vec{c}\\
\vec{a}.\vec{c^{\prime}}&\vec{b}.\vec{c^{\prime}}&1\\
\end{array}\right]^{-1}\left[\begin{array}{c}
\vec{V}.\vec{a^{\prime}}\\
\vec{V}.\vec{b^{\prime}}\\
\vec{V}.\vec{c^{\prime}}\end{array}\right]
=\left[
\begin{array}{ccc}
1 &0&0\\
0&1&0\\
0&0&1\\
\end{array}\right]^{-1}\left[\begin{array}{c}
\vec{V}.\vec{a^{\prime}}\\
\vec{V}.\vec{b^{\prime}}\\
\vec{V}.\vec{c^{\prime}}\end{array}\right].
\end{eqnarray}
In the light of equation (4) and equation (16), any arbitrary vector $\vec{V}$ is expanded in a non-orthogonal basis ($\vec{a}, \vec{b}, \vec{c}$) 
\begin{equation}
\vec{V} = (\vec{V}.\vec{a^{\prime}})\vec{a}+(\vec{V}.\vec{b^{\prime}})\vec{b}+(\vec{V}.\vec{c^{\prime}})\vec{c}
\end{equation}
where ($\vec{a^{\prime}}, \vec{b^{\prime}}, \vec{c^{\prime}}$) are reciprocal (dual) vectors.  Similar expansion of the vector $\vec{V}$ in the dual basis ($\vec{a^{\prime}}, \vec{b^{\prime}}, \vec{c^{\prime}}$) [1,2]
\begin{equation}
\vec{V} = (\vec{V}.\vec{a})\vec{a^{\prime}}+(\vec{V}.\vec{b})\vec{b^{\prime}}+(\vec{V}.\vec{c})\vec{c^{\prime}}.
\end{equation}
The coefficients of expansion in the basis ($\vec{a}, \vec{b}, \vec{c}$) as in equation (17) are called contravariant components of the vector $\vec{V}$ and the coefficients of expansion in the reciprocal(dual) basis ($\vec{a^{\prime}}, \vec{b^{\prime}}, \vec{c^{\prime}}$) of the same  vector  $\vec{V}$ as in equation (18) are called co-variant components [3]. Hence when we say a contravariant vector or covariant vector what we mean the component of a physical vector in two different non-orthogonal basis which are dual (reciprocal) to each other. An orthonormal basis is self-dual, there no distinction between contravariant and covariant component of a vector.

The expansion in equation (17) or in equation (18) similar to the orthogonal basis expansion in  equation (1) only change, the coefficients of expansion contain the vectors from (reciprocal) dual basis. Both equation (17) and equation (18) are contained in the completeness relation in the dyadic form
\begin{equation}
I=\vec{a}\vec{a^{\prime}}+\vec{b}\vec{b^{\prime}}+\vec{c}\vec{c^{\prime}}.
\end{equation}
Completeness relation in equation (19) is similar to completeness relation for orthonormal basis in equation (2) because of the notion contravariant and covariant vectors. We can state two set of non-orthogonal basis ($\vec{a},\vec{b},\vec{c}$) and ($\vec{a^{\prime}},\vec{b^{\prime}},\vec{c^{\prime}}$) which are reciprocal (dual) to each other behave like an orthonormal basis in unison. From equation (14), we can say $\vec{a^{\prime}}$ is effectively $\frac{1}{\vec{a}}$ . Pictorial visualisation of contravariant components and covariant components of a vector possible in two dimension [8,9].

In Quantum Mechanics the completeness relation for orthonormal eigen basis $\{|\phi_{i}\rangle\}$ of an Hermitian operator
\begin{equation}
I=\sum_{i} |\phi_{i}\rangle\langle\phi_{i}|.
\end{equation}
Equation (20) is analogous to an expansion of a vector $\vec{V}$ in (real) Euclidean space [6]. Equation (20) is the complex version of equation (2). We can also think of the two factors in equation (20) in two different spaces. The post-factor$|\phi_{i}\rangle$ a vector in Ket-space and the prefactor $\langle\phi_{i}|$  in Bra-space. Kets are analogues of contravariant vectors and Bras analogues of covariant vectors  [7].

Scalar product of two vectors $\vec{U}$ and $\vec{V}$ follow from the completeness relation in equation (19) 
\begin{equation}
\vec{U}.\vec{V}= (\vec{V}.\vec{a})(\vec{U}.\vec{a'}) + (\vec{V}.\vec{b})(\vec{U}.\vec{b'} )+ (\vec{V}.\vec{c})(\vec{U}.\vec{c'}).
\end{equation}
Under the change of basis, scalar product $\vec{U}.\vec{V}$ is invariant, $\vec{V}$ is contravariant and $\vec{U}$ is covariant. The equation (21) is  similar to inner product in Ket-Bra notation
\begin{equation}
\langle\Phi|\Psi\rangle=\sum_{i}\langle\Phi |\phi_{i}\rangle\langle\phi_{i}|\Psi\rangle.
\end{equation}

We introduce $\vec{a}=e_{1},\vec{b}=e_{2},\vec{c}=e_{3}$ for contravariant basis and $\vec{a^{\prime}} = e^{1},\vec{b^{\prime}} = e^{2},\vec{c^{\prime}} = e^{3} $ for covariant basis [4]. With this notation equation (15) and equation (19) become
\begin{equation}
I=e_{\mu}e^{\mu}
\end{equation}
\begin{equation}
e_{i}.e^{j}=\delta^{j}_{i}
\end{equation}
where summation over dummy indices is understood. $\delta^{j}_{i}$ is standard Kronecker delta function. With the introduction of superscript and subscript  notation we generalise equation (23) and equation (24) to n-dimensional Euclidean space. The contravariant component of any arbitrary vector $\vec{A}$ is  $A^{i}$ with superscript index  and covariant component is $A_{i}$ with subscript index are taken to be understood. The dimension of contravariant vector is  the inverse of the covariant vector and hence we expect the behaviour of contravariant vector and covariant vector under co-ordinate transformation inverse to each other. For a scalar function $\mathsf{f}$ in n-dimensional Euclidean space, we have from multivariate calculus
\begin{equation}
\mathsf{d} f = \frac{\partial f}{\partial x^{i}}d x^{i}.
\end{equation}
Under co-ordinate transformation  $\mathsf{df}$ is scalar invariant, the $\mathsf{dx^{i}}$ and $\mathsf\frac{\partial f}{\partial x^{i}}$ will transform inversely to compensate so that $\mathsf{df}$ remain invariant. We choose displacement vector $\mathsf(dx^{1}, dx^{2}, ...., dx^{n})$ as prototype contravariant vector and gradient of scalar function $\mathsf(\frac{\partial f}{\partial x^{1}},\frac{\partial f}{\partial x^{2}},...,\frac{\partial f}{\partial x^{n}})$ as prototype covariant vector. With this motivation we choose $\mathsf(dx^{1}, dx^{2}, ...., dx^{n})$ as covariant basis and $\mathsf(\frac{\partial }{\partial x^{1}},\frac{\partial }{\partial x^{2}},...,\frac{\partial }{\partial x^{n}})$ as contravariant basis in the co-ordinate representation and this notation is used in the relativistic theory.

Any contravariant vector $A^{i}\frac{\partial }{\partial x^{i}}$  under co-ordinate transformation behave like $\mathsf{dx^{i}}$
\begin{equation}
\mathsf{d} {x^{\prime}}^{i}= \frac{\partial {x^{\prime}}^{i}}{\partial x^{j}}d x^{j}
\end{equation}
\begin{equation}
\mathsf {A^{\prime}}^{i}= \frac{\partial {x^{\prime}}^{i}}{\partial x^{j}} A^{j}
\end{equation}
In equation (27) components of contravariant vector contra-vary with change of co-ordinate basis and hence the name contravariant vector.
Any covariant vector $A_{i}dx^{i}$  under co-ordinate transformation behave like $\mathsf\frac{\partial f}{\partial x^{i}}$
\begin{equation}
\mathsf \frac{\partial f}{\partial {x^{\prime}}^{i}}= \frac{\partial {x}^{j}}{\partial {x^{\prime}}^{i}} \frac{\partial f}{\partial {x}^{j}}
\end{equation}
\begin{equation}
\mathsf {A^{\prime}}_{i}= \frac{\partial {x}^{j}}{\partial {x^{\prime}}^{i}} A_{j}
\end{equation}
In equation (29) components of covariant vector (dual) co-vary with change of co-ordinate basis and hence the name covariant vector. In general direction vector like velocity vector is contravariant vector and dual vector like gradient (effectively like division by a vector) is a covariant vector, it is also called one-form.

The first fundamental form of a metric manifold
\begin{equation}
ds^2=g_{\mu\nu}(x)dx^{\mu}dx^{\nu}.
\end{equation}
The equation (30) can written as
\begin{equation}
I= \frac{dx_{\mu}}{ds}\frac{dx^{\mu}}{ds}.
\end{equation}
where $\frac{dx_{\mu}}{ds}=g_{\mu\nu}(x)\frac{dx^{\nu}}{ds}$ covariant component and  $\frac{dx^{\mu}}{ds}$ contravariant component of tangent vectors at a point on the manifold and  we have $g_{\mu\nu}=e_{\mu}.e_{\nu}$ and $g^{\mu\nu}=e^{\mu}.e^{\nu}$ [4]. Equation (31) is a completeness relation like equation (23). The contravariant tensor conjugate to $g_{ij}$ is denoted by $g^{ij}$ satisfy [5]
\begin{equation}
g_{ij}g^{ik}=\delta^{k}_{j}.
\end{equation}
Equation (32) the non-Euclidean geometrical version of the notion contravariant and covariant vectors and it can obtained from equation (30). Because the components of vectors are contravariant and those of covectors are covariant, the vectors themselves are often referred to as being contravariant and the covectors as covariant. This usage is not universal, however, since vectors push forward - are covariant under deffeomorphism - and covectors pull back - are contravariant under deffeomorphism [10]. In metric manifold we convert contravariant to covariant and vice-versa with metric tensor but in manifold without metric contravariant and covariant vectors are totally different entities.

\vspace{0.5cm}

{\noindent{\bf{Acknowledgements}}}

\vspace{0.5cm}

I thank Director, IISER, Bhopal  for providing constant help and encouragement during the completion of this work. The PhD students of Physics Department, IISER, Bhopal is acknowledged with thanks for their discussion during the course of  Electrodynamics.

\newpage 

\vspace{0.5cm}

{\noindent{\bf{Appendix}}}

\vspace{0.5cm}

We are given the matrix 
\begin{eqnarray}
A=\left[
\begin{array}{ccc}
1 &\vec{a}.\vec{b}&\vec{a}.\vec{c}\\
\vec{a}.\vec{b}&1&\vec{b}.\vec{c}\\
\vec{a}.\vec{c}&\vec{b}.\vec{c}&1\\
\end{array}\right]^{-1}
\end{eqnarray}
$A$ is a symmetric matrix, as it is a inverse of a symmetric matrix.
\begin{eqnarray}
A=\left[
\begin{array}{ccc}
A_{11}&A_{12}&A_{13}\\
A_{21}&A_{22}&A_{23}\\
A_{13}&A_{23}&A_{33}
\end{array}\right].
\end{eqnarray}
\begin{equation}
\left.
\begin{array}{ccc}
A_{11}=\frac{1}{D}[1-(\vec{b}.\vec{c})^2]\\
A_{22}=\frac{1}{D}[1-(\vec{a}.\vec{c})^2]\\
A_{33}=\frac{1}{D}[1-(\vec{a}.\vec{b})^2]\\
A_{12}=\frac{1}{D}[(\vec{a}.\vec{c})(\vec{b}.\vec{c})-(\vec{a}.\vec{b})]\\
A_{13}=\frac{1}{D}[(\vec{a}.\vec{c})(\vec{b}.\vec{c})-(\vec{a}.\vec{b})]\\
A_{23}=\frac{1}{D}[(\vec{a}.\vec{b})(\vec{a}.\vec{c})-(\vec{b}.\vec{c})]\\
A_{21}=\frac{1}{D}[(\vec{a}.\vec{c})(\vec{b}.\vec{c})-(\vec{a}.\vec{b})]\\
A_{31}=\frac{1}{D}[(\vec{a}.\vec{c})(\vec{b}.\vec{c})-(\vec{a}.\vec{b})]\\
A_{32}=\frac{1}{D}[(\vec{a}.\vec{b})(\vec{a}.\vec{c})-(\vec{b}.\vec{c})]\\
\end{array}
\right\}
\end{equation}
where determinant $D$ is given by equation (10).

\vspace{0.5cm}

\newpage

{\noindent{\bf{References}}}

\vspace{0.5cm}

\begin{enumerate}
\item Murray R. Spiegel,\, Theory and Problems of Vector Analysis (SI metric edition),\,Schaum's Outline Series,\,(1974).
\item George B.\, Arfken and  Hans J.\, Weber,\, Mathematical Methods for Physicists (sixth edition),\,Academic Press,\,(2005).
\item Barry Spain,\, Vector Analysis (third edition),\,John Wiley and Sons,\,(1960).
\item A.\,I.\,Borisenko and I.\,E.\,Tarapov,\,Vector and Tensor Analysis with Applications,\,Dover Publications,\,(1979).
\item Barry Spain,\, Tensor Calculus (first edition),\,Dover Publications,\,(2003).
\item J.\,J.\,Sakurai,\,Modern Quantum Mechanics (revised edition),\, Pearson Education,\,(1994).
\item Eugen Merzbacker,\,Quantum Mechanics (third edition),\,John Wiley \& sons,\,(1999).
\item S.\,R.\,Deans,\, Covariant and Contravariant Vectors,\,Mathematica Magazine, Vol. {\bf{44}}, MAA (1971).
\item Dwight\,E.\,Neuenschwander,\,``Question 55. Are there pictorial examples that distinguish covariant and contravariant vectors?,''\\ Am.\,J.\,Phys.\,{\bf{65(1)}},\,11\,(1997).
\item ``http://en.wikipedia.org/wiki/Covariance-and-contravariance.''
\end{enumerate}
\end{document}